\documentclass[aip,jap,amsmath,amssymb,preprint]{revtex4-1}

\usepackage{graphicx}
\usepackage{dcolumn}
\usepackage{bm}
\usepackage{braket}

\begin{document}


\title{First-principles simulation of the optical response of bulk and thin-film $\alpha$-quartz irradiated with an ultrashort intense laser pulse}

\author{Kyung-Min Lee}
 \affiliation{Advanced Photonics Research Institute, Gwangju Institute of Science and Technology, Gwangju 500-712, Republic of Korea}

\author{Chul Min Kim}
 \affiliation{Advanced Photonics Research Institute, Gwangju Institute of Science and Technology, Gwangju 500-712, Republic of Korea}
 \affiliation{Center for Relativistic Laser Science, Institute for Basic Science, Gwangju 500-712, Republic of Korea}

\author{Shunsuke A. Sato}
 \affiliation{Graduate School of Pure and Applied Sciences, University of Tsukuba, Tsukuba 305-8571, Japan}

\author{Tomohito Otobe}
 \affiliation{Kansai Photon Science Institute, Japan Atomic Energy Agency, Kizugawa, Kyoto 619-0215, Japan}

\author{Yasushi Shinohara}
 \affiliation{Graduate School of Pure and Applied Sciences, University of Tsukuba, Tsukuba 305-8571, Japan}
 \affiliation{Max-Planck-Institut f\"{u}r Mikrostrukturphysik, Weinberg 2, D-06120 Halle, Germany}

\author{Kazuhiro Yabana}
 \affiliation{Graduate School of Pure and Applied Sciences, University of Tsukuba, Tsukuba 305-8571, Japan}
 \affiliation{Center for Computational Sciences, University of Tsukuba, Tsukuba 305-8571, Japan}

\author{Tae Moon Jeong}
 \affiliation{Advanced Photonics Research Institute, Gwangju Institute of Science and Technology, Gwangju 500-712, Republic of Korea}
 \affiliation{Center for Relativistic Laser Science, Institute for Basic Science, Gwangju 500-712, Republic of Korea}
 \email{jeongtm@gist.ac.kr}

\begin{abstract}
A computational method based on a first-principles multiscale simulation has been used for calculating the optical response and the ablation threshold of an optical material irradiated with an ultrashort intense laser pulse. The method employs Maxwell's equations to describe laser pulse propagation and time-dependent density functional theory to describe the generation of conduction band electrons in an optical medium. Optical properties, such as reflectance and absorption, were investigated for laser intensities in the range $10^{10} \, \mathrm{W/cm^{2}}$ to $2 \times 10^{15} \, \mathrm{W/cm^{2}}$ based on the theory of generation and spatial distribution of the conduction band electrons. The method was applied to investigate the changes in the optical reflectance of $\alpha$-quartz bulk, half-wavelength thin-film and quarter-wavelength thin-film and to estimate their ablation thresholds. Despite the adiabatic local density approximation used in calculating the exchange--correlation potential, the reflectance and the ablation threshold obtained from our method agree well with the previous theoretical and experimental results. The method can be applied to estimate the ablation thresholds for optical materials in general. The ablation threshold data can be used to design ultra-broadband high-damage-threshold coating structures.
\end{abstract}


\maketitle

\section{\label{sec:intro} Introduction}

The advances made in femtosecond (fs) high-power laser technology in the last decade have made it possible to achieve laser intensities as high as $10^{22} \, \mathrm{W/cm^{2}}$ \cite{Bahk:2004}. With such a wide range of laser intensities available for investigations, the optical response of a material can be expected to show fairly different characteristics with varying intensity. For example, at very low intensities below $10^{10} \, \mathrm{W/cm^{2}}$, the optical properties of a medium follow a linear response to laser intensity variation \cite{Hecht:2001, Born:1999}, but start showing a nonlinear response as the laser intensity increases beyond a certain level \cite{Boyd:2008}. However, at still higher laser intensities of greater than $10^{14} \, \mathrm{W/cm^{2}}$, the optical medium suddenly starts behaving like a plasma medium, and its optical properties follow the properties of a plasma medium \cite{Kruer:1989}. In the intermediate intensity range ($10^{11} \, \mathrm{W/cm^{2}}$ to $10^{14} \, \mathrm{W/cm^{2}}$), the physical behavior of an optical medium is very complicated and many interesting phenomena, e.g., generation and heating of conduction band (CB) electrons and energy transfer to the lattice, followed by melting, boiling and ablation of the material, can be observed. These behaviors are related to the transition mechanism from solid to plasma and have been intensively studied in previous reports. A theoretical understanding of the laser--matter interactions in the intermediate intensity range is, therefore, of great interest. In addition, it can also provide important insights into laser-induced damage and ablation of optical materials in general.

Studies on laser-induced damage date back to as far as the late 1960s. The dependence of the damage on laser characteristics such as the wavelength, pulse duration and energy fluence as well on material type was investigated by Wood using nanosecond (ns) laser pulses \cite{Wood:2003}. Later, investigations of laser-induced damage in the picosecond (ps) and fs regime gained significance when the advent of the chirped-pulse amplification (CPA) technique \cite{Strickland:1985} made it feasible to develop fs and petawatt-class laser systems \cite{Sung:2010,Yu:2012}. In particular, laser ablation occurring on the fs time scale became critical because a laser pulse duration of few tens of fs is much shorter than the time scale for electron energy transfer to the lattice and subsequent lattice heating. In 1995, Stuart $\mathit{et \, al.}$, investigated the laser-induced damage threshold at $1053 \, \mathrm{nm}$ and $526 \, \mathrm{nm}$ for pulse durations ranging from $270 \, \mathrm{fs}$ to $1 \, \mathrm{ns}$, through a theoretical model based on CB electron production via multiphoton ionization, Joule heating and collisional ionization \cite{Stuart:1995}. Subsequent studies by other groups were conducted for a more accurate analysis of the damage and ablation threshold by including energy dependence of the CB electrons \cite{Rethfeld:2002} and nonlinear pulse propagation effect in a medium \cite{Penano:2005, Petrov:2008, Gulley:2010, Apalkov:2012} in the fs regime. However, all these studies were based on theoretical models that used experimental and/or empirical values of the material parameters such as ionization rate, refractive index, relaxation rate, and band structure. Hence, the need for developing a method that uses non-empirical values of the material parameters grew continuously in the search for a comprehensive and reliable method of investigating laser--matter interactions in the intermediate laser intensity range.

In this paper, we employ an alternative method to compute the optical response and the ablation threshold of an optical medium. In contrast to the previous studies, our method is based on first-principles simulations computed from fundamental equations. A multiscale approach using the wave equation and the time-dependent density functional theory (TDDFT) is applied to calculate directly the density of the CB electrons generated in the optical medium. The first report on the use of such a multiscale approach for investigation of the interaction between a laser pulse and an optical medium was made for crystalline silicon, where it was said to yield reliable results \cite{Yabana:2012}. In this approach, no empirical parameters and approximations were used except for information on the crystal structure and on the exchange--correlation potential. As far as these parameters and approximations are valid for a given set of conditions, our first-principles simulations can produce the most reliable and comprehensive results.

We applied the method to calculate the reflectance, the CB electron density and the absorbed energy for investigating the changes in the optical properties of bulk and thin-film $\alpha$-quartz (having different thicknesses) on being irradiated by fs laser pulses in the intensity range of $10^{10} \, \mathrm{W/cm^{2}}$ to $2 \times 10^{15} \, \mathrm{W/cm^{2}}$. By comparing the absorbed energy based on some criterion for laser-induced ablation, the ablation threshold can be computationally determined without the help of empirical values. The proposed approach can be easily applied to other optical materials and structures to design high-performance optical coatings, such as a high-damage-threshold broadband optical coating. The organization of the paper is as follows. Section \ref{sec:method} describes in brief the theoretical methods and the simulation details. The calculated results and discussion are presented in Section \ref{sec:results}. Finally, the conclusion of the paper is given in Section \ref{sec:conclusion}.

\section{\label{sec:method} Theoretical methods}

\subsection{\label{ssec:multiscale} Multiscale description of laser-matter interaction}

We employ a theoretical method and a computational code developed by some of the present authors \cite{Yabana:2012}. In the following, we briefly describe the formalism. The interaction between a laser pulse and matter involves two characteristic lengths: the wavelength of the laser pulse and the electronic structure size of the atoms constituting the matter. In the case of fs laser pulses, the former lies on the macroscopic scale comprising the $\mu\mathrm{m}$ range, while the latter lies on the microscopic scale comprising the $\mathrm{nm}$ range. Any first-principles description of the interaction should incorporate these two different scales simultaneously. Let \textbf{R} denote the macroscopic scale in which the laser pulse evolves and \textbf{r} the microscopic scale in which the electrons move. To describe the dynamics of electrons in a unit cell under an external electromagnetic field, the time-dependent Kohn--Sham (TDKS) equation is used \cite{Runge:1984}:
\begin{eqnarray}
\label{eq:tdks}
\mathrm{i}\hbar\frac{\partial}{\partial t}\psi_{i,\mathbf{R}}(\vec{r},t)&=&\biggl\{ \frac{1}{2m_{\mathrm{e}}}\left(-\mathrm{i}\hbar\nabla_{\mathbf{r}}+\frac{e}{c}\vec{A}_{\mathbf{R}}(t)\right)^{2}+V_{\mathrm{ion},\mathbf{R}}(\vec{r}) \nonumber \\
&&+V_{\mathrm{h},\mathbf{R}}(\vec{r},t)+V_{\mathrm{xc},\mathbf{R}}(\vec{r},t)\biggr\}\psi_{i,\mathbf{R}}(\vec{r},t),
\end{eqnarray}
where $\psi_{i,\mathbf{R}}$ is the $i$th Kohn--Sham (KS) orbital, $\vec{A}_{\mathbf{R}}$ the vector potential of the laser pulse in the Coulomb gauge, $V_{\mathrm{ion},\mathbf{R}}$ the ionic potential, $V_{\mathrm{h},\mathbf{R}}$ the Hartree potential and $V_{\mathrm{xc},\mathbf{R}}$ the exchange--correlation potential.

Since the laser pulse we considered slowly varies over the electronic length scale, it can be assumed that $\vec{A}_{\mathbf{R}}$ does not depend on $\vec{r}$. Once the TDKS equation is solved with a given vector potential, the electron density ($n_{\mathbf{R}}$) and current ($\vec{j}_{\mathbf{R}}$) can be calculated from the KS orbitals:
\begin{equation}
\label{eq:density}
n_{\mathbf{R}}(\vec{r},t)=\sum_{i}\left|\psi_{i,\mathbf{R}}(\vec{r},t)\right|^{2},
\end{equation}
\begin{eqnarray}
\label{eq:current}
\vec{j}_{\mathbf{R}}(\vec{r},t)&=&\frac{1}{2m_{\mathrm{e}}}\sum_{i}\biggl\{\psi_{i,\mathbf{R}}^{\ast}(\vec{r},t)\left(-\mathrm{i}\hbar\nabla_{\mathbf{r}}+\frac{e}{c}\vec{A}_{\mathbf{R}}(t)\right)\psi_{i,\mathbf{R}}(\vec{r},t) \nonumber \\
&& -\psi_{i,\mathbf{R}}(\vec{r},t)\left(-\mathrm{i}\hbar\nabla_{\mathbf{r}}-\frac{e}{c}\vec{A}_{\mathbf{R}}(t)\right)\psi_{i,\mathbf{R}}^{\ast}(\vec{r},t)\biggr\}.
\end{eqnarray}
This microscopic current is averaged over a unit cell to define the macroscopic current ($\vec{J}_{\mathbf{R}}$) as:
\begin{equation}
\label{eq:maccurrent}
\vec{J}_{\mathbf{R}}(t)=\frac{1}{\Omega}\int_{\Omega} \vec{j}_{\mathbf{R}}(\vec{r},t) \, \mathrm{d}\vec{r},
\end{equation}
where $\Omega$ is the unit cell volume. It should be noted that there is also a contribution to the current from a nonlocal pseudopotential. The propagation of the laser pulse is described by the wave equation with the macroscopic current as its source term:
\begin{equation}
\label{eq:we}
\frac{1}{c^2}\frac{\partial^2}{\partial t^2}\vec{A}_{\mathbf{R}}(t)-\nabla^2_{\mathbf{R}}\vec{A}_{\mathbf{R}}(t)=-\frac{4\pi e}{c}\vec{J}_{\mathbf{R}}(t).
\end{equation}
The vector potential obtained by solving Eq. (\ref{eq:we}) is used to solve Eq. (\ref{eq:tdks}) at the next time step. The interaction between a laser pulse and matter can be fully described by solving Eqs. (\ref{eq:tdks}) and (\ref{eq:we}) self-consistently via the macroscopic current and the vector potential \cite{Yabana:2012}. It should be noted that the electron motion is restricted to be within a unit cell; non-local processes such as electron transport among unit cells cannot be accounted for in the present method. Moreover, the ionic motion is neglected since the motion of ions is slow enough in comparison with electrons due to their large mass. However, these are beyond the scope of our interest since we consider laser intensities smaller than $10^{17} \, \mathrm{W/cm^{2}}$ and wavelength of the pulse in the near-visible region.

\subsection{\label{ssec:details} Simulation details}

For the sake of simplicity of the laser--matter interaction geometry, the case of normal incidence of the pulse was considered in the simulation. The $\left(\bar{2}10\right)$ surface of $\alpha$-quartz was taken to be the transverse plane. The laser pulse was assumed to be linearly polarized along the $z$-axis and propagating along the $x$-axis. It had a wavelength of $\lambda_{0}=800 \, \mathrm{nm}$, corresponding to a photon energy of $\hbar\omega=1.55 \, \mathrm{eV}$ and a pulse duration of $T=20 \, \mathrm{fs}$. In the simulation, a uniform laser intensity was assumed in the transverse plane. The vector potential of the incident pulse was given as
\begin{equation}
A_{X}(t)=-\frac{E_{0}}{\omega}\sin^{2}\left\{ \frac{\pi\left(X-ct\right)}{cT}\right\} \cos\left\{ \frac{\omega\left(X-ct\right)}{c}\right\} 
\end{equation}
for $0<X-ct<cT$ and $A_{X}(t)=0$ otherwise. Here, $X$ denotes the macroscopic coordinate in the laser propagation direction and $E_{0}$ is the maximum electric field strength, which is related to the laser intensity ($I_{0}$) as $I_{0}=cE_{0}^{2}/8\pi$. The spatial step size along the $z$-axis had a value of $\Delta X=12.67 \, \mathrm{nm}$. The incident electric field was related to the vector potential by $E_{X}(t)=-\left(1/c\right)\cdot\left(dA_{X}(t)/dt\right)$.

The thickness of the $\alpha$-quartz sample was appropriately chosen so as to simulate bulk and thin-film structures. The $\alpha$-quartz bulk sample had a thickness of $d_{\mathrm{bulk}}=3.548 \, \mu\mathrm{m}$, which was considered large enough for the assumption that there was no reflection from the rear surface during pulse propagation. The thin-film samples of $\alpha$-quartz had thicknesses of $d_{\mathrm{HWTF}}=\lambda_{0}/2n_{0}=253.28 \, \mathrm{nm}$ for the half-wavelength thin film (HWTF) and $d_{\mathrm{QWTF}}=\lambda_{0}/4n_{0}=126.64 \, \mathrm{nm}$ for the quarter-wavelength thin film (QWTF). Here, $n_{0}$ is the refractive index of $\alpha$-quartz at $\lambda_{0}=800 \, \mathrm{nm}$. A value of $n_{0}=1.578$ obtained from the TDDFT calculations \cite{Otobe:2009} was used rather than an experimental value of $n_{0}=1.538$ \cite{Ghosh:1999}.

An orthogonal unit cell containing six $\mathrm{SiO_2}$ molecular units was used as a unit cell for $\alpha$-quartz. Three sides of the unit cell had lengths of $\mathrm{a}=9.28 \, \mathrm{a.u.}$, $\mathrm{b}=16.08 \, \mathrm{a.u.}$ and $\mathrm{c}=10.21 \, \mathrm{a.u.}$, respectively \cite{Schober:1993, Adeagbo:2008}. The calculation results numerically converged well when the sides were discretized into $26$, $36$ and $50$ points, respectively. The number of $k$-points in the reciprocal space was $4^{3}$ in the simulation.

The TDKS equations were solved by applying the explicit time evolution operator containing up to the fourth-order term in the Taylor expansion of the complete time evolution operator \cite{Yabana:1996, Yabana:2006}:
\begin{equation}
e^{-\mathrm{i}H_{\mathrm{KS}}\Delta t/\hbar}\approx\sum_{n=0}^{4}\frac{\left(-\mathrm{i} H_{\mathrm{KS}} \Delta t / \hbar \right)^{n}}{n!},
\end{equation}
where $\Delta t=0.2 \, \mathrm{a.u.}$ ($1 \, \mathrm{a.u.}$ of time corresponds to $0.024189 \, \mathrm{fs}$) is the time step and $H_{\mathrm{KS}}$ is the Kohn--Sham Hamiltonian in Eq. (\ref{eq:tdks}).

In this paper, we employed a norm-conserving pseudopotential \cite{Troullier:1991} with the separable approximation \cite{Kleinman:1982}. In this approximation, only the valence band electrons were explicitly treated in the simulation, while the effect of core electrons was included in the pseudopotential.

To calculate the exchange--correlation potential in Eq. (\ref{eq:tdks}), the adiabatic local density approximation (LDA) was used \cite{Perdew:1981}. The calculated band gap energy was $6.5 \, \mathrm{eV}$ in our simulation, while the experimental one was $9.0 \, \mathrm{eV}$ \cite{Arnold:1994}. This underestimation of the band gap energy is a well-known characteristic of the LDA. Consequently, the number of photons responsible for interband transitions is reduced, which indicates that more CB electrons are generated at the same laser intensity. This discrepancy in the band gap energy should be kept in mind when the calculated quantities are compared to the experimental ones. A more quantitative evaluation of the band gap energy can be systematically achieved by using an elaborate functional, e.g., meta-GGA \cite{Tran:2009, Rasanen:2010}, which is being implanted in our simulation code, and will be presented in a further study.

\section{\label{sec:results} Results and discussion}

This section consists of three major parts. The first part describes the laser intensity dependence of the reflectance for bulk and thin-film samples of $\alpha$-quartz irradiated by an ultrashort intense laser pulse (see \ref{ssec:reflectance}). The second part describes explicitly the generation and spatial distribution of the CB electrons, which are responsible for the change in the reflectance, in the $\alpha$-quartz medium (see \ref{ssec:CBE} and \ref{ssec:thickness}). In the last part, the extent of laser-induced ablation is estimated based on the energy absorbed by the CB electrons in the medium (see \ref{ssec:absorption}).

\subsection{\label{ssec:reflectance} Reflectance as a function of laser intensity}

\begin{figure}
\includegraphics[width=80mm]{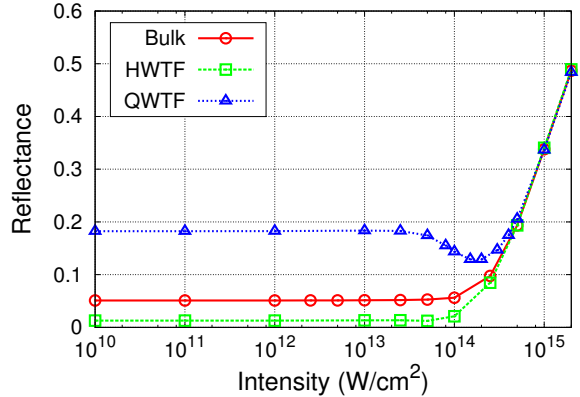}
\caption{\label{fig:reflectance} Reflectance of $\alpha$-quartz as a function of laser intensity: bulk (circles and solid line), HWTF (squares and dashed line) and QWTF (triangles and dotted line).}
\end{figure}

The optical response of the $\alpha$-quartz materials under investigation was described by the reflectance curve for various laser intensity conditions. The reflectance in our simulation was calculated as the fraction of power of the incident laser pulse that is reflected at the surface when the reflected and the transmitted pulses are well separated. In this study, only static results, which mean the results after pulse propagation is over, were considered, although our simulation could intrinsically deal with time-dependent processes. Figure \ref{fig:reflectance} shows the calculated reflectances of $\alpha$-quartz materials under laser intensities of $10^{10} \, \mathrm{W/cm^{2}}$ to $2 \times 10^{15} \, \mathrm{W/cm^{2}}$. As shown in Fig. \ref{fig:reflectance}, at low intensities below $2.5 \times 10^{13} \, \mathrm{W/cm^{2}}$, all the reflectances have different but constant values. The constant values of the reflectances at low intensities can be attributed to the linear response of lossless dielectric materials. According to Fresnel's equation \cite{Born:1999}, the reflectance of a bulk material is given by
\begin{equation}
\label{eq:Rbulk}
R_{\mathrm{bulk}}=\left(\frac{n_0-1}{n_0+1}\right)^2,
\end{equation}
where $n_{0}$ refers to the refractive index of the material at a given wavelength. Eq. (\ref{eq:Rbulk}) yields $R_{\mathrm{bulk}}=0.050$ with $n_{0}=1.578$, which is almost the same as $R_{\mathrm{bulk}}=0.051$ obtained from the simulation (see Fig. \ref{fig:reflectance}). For a thin film, interference between the secondary waves reflected from the front and rear surfaces should be considered as well. The formula is thus modified to \cite{Dressel:2002}
\begin{equation}
\label{eq:Rfilm}
R_{\mathrm{film}}(\beta)=\frac{4R_{\mathrm{bulk}}\sin^{2}\beta}{(1-R_{\mathrm{bulk}})^{2}+4R_{\mathrm{bulk}}\sin^{2}\beta},
\end{equation}
where $\beta=2\pi n_{0}d/\lambda_{0}$ and $d$ is the thickness of the film. Eq. (\ref{eq:Rfilm}) gives $R_{\mathrm{HWTF}}(\pi)=0$ and $R_{\mathrm{QWTF}}(\pi/2) =0.181$, which are almost identical to the results of $R_{\mathrm{HWTF}}(\pi)=0.013$ and $R_{\mathrm{QWTF}}(\pi/2)=0.184$ obtained from Fig. \ref{fig:reflectance}. The slight differences might have been caused by material dispersion because we used a $20 \, \mathrm{fs}$ pulse. Material dispersion cannot be described by a single Fresnel's equation at a fixed frequency. In the case of thin films, interference between the secondary waves generated at the front and rear surfaces plays a dominant role in determining the reflectances.

It should be noted that secondary waves can also be generated inside the medium, but they get summed to zero along the reflection direction as long as the medium has a high uniform density \cite{Hecht:2001}, according to the Ewald--Oseen extinction theorem \cite{Fearn:1996}. However, as the laser intensity increases, the CB electrons, which behave like free electrons, can be non-uniformly generated inside the medium. By absorbing and reflecting the laser pulse, the non-uniform distribution of the CB electrons can change the interference pattern among the secondary waves and may result in a change in reflectance. As shown in Fig. \ref{fig:reflectance}, as the laser intensity is increased beyond $2.5 \times 10^{13} \, \mathrm{W/cm^{2}}$, the reflectance for QWTF and HWTF first decreases--though this decrease is only $8 \, \%$ for HWTF at an intensity of $5 \times 10^{13} \, \mathrm{W/cm^{2}}$--and thereafter starts increasing. In constrast, the reflectance for bulk $\alpha$-quartz is almost constant up to $5 \times 10^{13} \, \mathrm{W/cm^{2}}$ and then increases monotonously. To understand the role of CB electrons in changing the reflectance, the generation and spatial distribution of the CB electrons should be analyzed in detail, as further discussed in Sect. \ref{ssec:CBE} and Sect. \ref{ssec:thickness}.

At high intensities, $I_{0}>5 \times 10^{14} \, \mathrm{W/cm^{2}}$, the reflectances for all the structures rapidly increased and converged to the same value. Convergence in the reflectances implies that the CB electrons, which are mostly generated at the front surface, play a leading role in the bringing about changes in reflectance. To investigate this effect quantitatively, we define a parameter called the skin depth \cite{Gibbon:2005}, which expresses light penetration into the medium as
\begin{equation}
\label{eq:skindepth}
l_{\mathrm{s}}=\frac{c}{\sqrt{\omega_{\mathrm{p}}^{2}-\omega^{2}}},
\end{equation}
where $c$ is the velocity of light, $\omega_{\mathrm{p}}=\sqrt{4\pi e^{2} N_{\mathrm{CB}}/m_{\mathrm{e}}^{\ast}}$ the plasma frequency and $\omega$ the laser frequency. With the CB electron density ($N_{\mathrm{CB}}$) calculated at an intensity of $5 \times 10^{14} \, \mathrm{W/cm^{2}}$, we estimated the skin depth ($l_{\mathrm{s}}$) as $28 \, \mathrm{nm}$ for HWTF, $30 \, \mathrm{nm}$ for QWTF and $29 \, \mathrm{nm}$ for bulk, depths that are much smaller than the thickness of even QWTF. At intensities higher than $5 \times 10^{14} \, \mathrm{W/cm^{2}}$, the skin depths became much smaller due to the high $N_{\mathrm{CB}}$. This small skin depth indicates that the transmitted waves barely reached the rear surface and the reflection mainly occurred at the front surface by the many CB electrons present there.

It should be noted that the optical Kerr effect can also contribute to the change in reflectance. The third-order nonlinear susceptibility of $\alpha$-quartz is given as $\chi^{(3)}=3.81 \times 10^{-14} \, \mathrm{esu}$ \cite{Buchalter:1982} and the corresponding nonlinear refractive index is $n_{2}=6.04 \times 10^{-16} \, \mathrm{cm^{2}/W}$ with $n_{0}=1.578$. Substituting this value of $n_{0}$ into $n=n_{0}+n_{2}I_{0}$ in Eq. (\ref{eq:Rbulk}) at an intensity of $I_{0}=10^{15} \, \mathrm{W/cm^{2}}$, the reflectance for bulk $\alpha$-quartz is calculated to be $0.14$, which is less than $0.34$ obtained from the simulation. Therefore, the optical Kerr effect would be less important for the change of the bulk reflectance, nor is it valid for the case of thin-film reflectance because interference among the secondary waves has been ignored.

\subsection{\label{ssec:CBE} Generation of CB electrons}

\begin{figure}
\includegraphics[width=80mm]{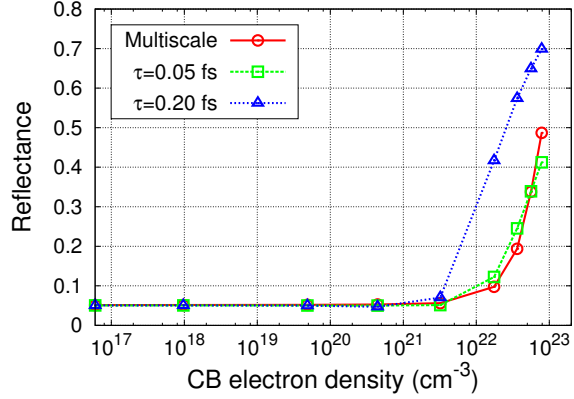}
\caption{\label{fig:drude} Change in the bulk reflectance calculated from Eq. (\ref{eq:drude}) as a function of the calculated CB electron density ($N_{\mathrm{CB}}$) and the collision time ($\tau$). The open circles and the solid line represent our simulation results and the open rectangles and the dashed line represent the best fit (from Eq. (\ref{eq:drude})) to our simulation results, keeping the effective electron mass fixed at $m_{\mathrm{e}}^{\ast}=0.5 \, m_{\mathrm{e}}$. The open triangles and the dotted line represent the fit from the experimental value of $\tau=0.2 \, \mathrm{fs}$.}
\end{figure}

In Section \ref{ssec:reflectance}, changes in the reflectance were attributed to the generation of CB electrons. For large band gap materials under a weak and infrared laser pulse, the valence band (VB) electrons cannot be directly excited into the CB since the photon energy is smaller than the band gap energy. As the laser intensity increases, material begins to absorb multiple photons and then the VB electrons can be excited to the CB. In our simulation, the increase in the CB electrons was proportional to the laser intensity ($I_{0}$) in the form of $I_{0}^{4}$. This indicates that the excitation occurred by means of a four-photon absorption process that overcame the calculated band gap energy of $6.3 \, \mathrm{eV}$. The generated CB electrons absorbed laser energy through the inverse Bremsstrahlung process, which resulted in the change in optical response of material such as permittivity. To confirm this scenario, we calculated the reflectance by using a modified Drude model as an empirical model, which includes free electron generation and effect on decrease of VB electrons by the electron excitation to CB, and compared the calculated results with our simulation results. According to the model, the permittivity of $\alpha$-quartz can be written as
\begin{eqnarray}
\label{eq:drude}
\epsilon&=&1+\frac{N_{\mathrm{VB}}}{N_{\mathrm{VB}}^{0}}\left(\epsilon_{0}-1\right)-\frac{\omega_{\mathrm{p}}^{2}}{\omega^{2}\left(1+i\tilde{\nu}\right)} \nonumber \\
&=&\epsilon_{0}-N_{\mathrm{CB}}\cdot\left(\frac{1}{N_{\mathrm{cr}}\cdot\left(1+\mathrm{i}\tilde{\nu}\right)}+\frac{\epsilon_{0}-1}{N_{\mathrm{VB}}^{0}}\right),
\end{eqnarray}
where $N_{\mathrm{VB}}^{0}$ is the number of the initial VB electrons, $N_{\mathrm{VB}}^{0}=N_{\mathrm{VB}}+N_{\mathrm{CB}}=4.25 \times 10^{23} \, \mathrm{cm}^{-3}$, $\tilde{\nu}$ is the relative collision frequency given by $\tilde{\nu}=\nu/\omega$ and $N_{\mathrm{cr}}$ is the critical density by $N_{\mathrm{cr}}=\omega^2 m_{\mathrm{e}}^{\ast}/4\pi e^{2}=8.7 \times 10^{20} \, \mathrm{cm}^{-3}$. A collision time ($\tau$) and an effective mass of $\alpha$-quartz ($m_{\mathrm{e}}^{\ast}$) were defined as $\tau=1/\nu$ and $m_{\mathrm{e}}^{\ast}=0.5 m_{\mathrm{e}}$ \cite{Vexler:2005}, respectively. In the simulation, the CB electron density was calculated at the macroscopic points, i.e., the unit cells, in the medium as follows:
\begin{equation}
\label{eq:NCB}
N_{\mathrm{CB},\mathbf{R}}(t)=\sum_{i,j}\left(\delta_{ij}-\left| \braket{\psi_{i,\mathbf{R}}(t=0)|\psi_{j,\mathbf{R}}(t)} \right|^2\right),
\end{equation}
where $i$ and $j$ are indices for the Kohn--Sham orbitals and $\psi_{i,\mathrm{R}}(t=0)$ is the ground-state Kohn--Sham orbital.

Figure \ref{fig:drude} shows the change in the reflectance of bulk $\alpha$-quartz calculated from Eq. (\ref{eq:drude}) as a function of the calculated $N_{\mathrm{CB}}$ and $\tau$. It should be noted that the calculated $N_{\mathrm{CB}}$ was taken as the value of Eq. (\ref{eq:NCB}) after the pulse passed the medium. The calculation based on the model could qualitatively reproduce our simulation results; the bulk reflectance increased with increase in the CB electrons. For more quantitative comparison, the best fit to the simulation result was achieved when $\tau=0.05 \, \mathrm{fs}$, which was much smaller than the experimental value, $\tau=0.2 \, \mathrm{fs}$ \cite{Mao:2003}. The discrepancy could be understood by the fact that our simulation intrinsically considers the time-dependent CB electrons, which means that when the pulse reaches to medium, CB electrons are initially zero and increase with propagation of the pulse in the medium. However, in Eq. (\ref{eq:drude}), only static values of CB electrons, which are the values after the pulse passes the medium, are considered. Therefore the quantitative comparison between the model and our simulation could be difficult.

It should also be mentioned that the underestimation of the band gap energy in our simulation might affect the results. Four-photon absorption rather than six-photon absorption, which is the correct one for an experimental band gap energy of about $9 \, \mathrm{eV}$, can generate more CB electrons at the same laser intensity. This might cause an increase in the reflectance at a lower intensity than that expected from six-photon absorption. When more elaborate exchange--correlation functionals reproducing six-photon absorption are used, the increase in reflectance would occur at higher intensities.

\subsection{\label{ssec:thickness} Dependence of CB electron generation on material thickness}

The empirical model described by Eq. (\ref{eq:drude}) could qualitatively explain the changes in reflectance of bulk $\alpha$-quartz based on generation of CB electrons. However, it may not be a suitable explanation for the case of thin films because the interference effect was not considered in the model. As mentioned earlier, the CB electrons generated inside the medium absorb and reflect the laser pulse, resulting in a change in the condition for interference, which plays a dominant role in determining the reflectance of thin films. In this regard, it is important to know the spatial distribution of the CB electrons for understanding how it changes the interference condition, and hence the reflectance.

\begin{figure*}
\includegraphics[width=120mm]{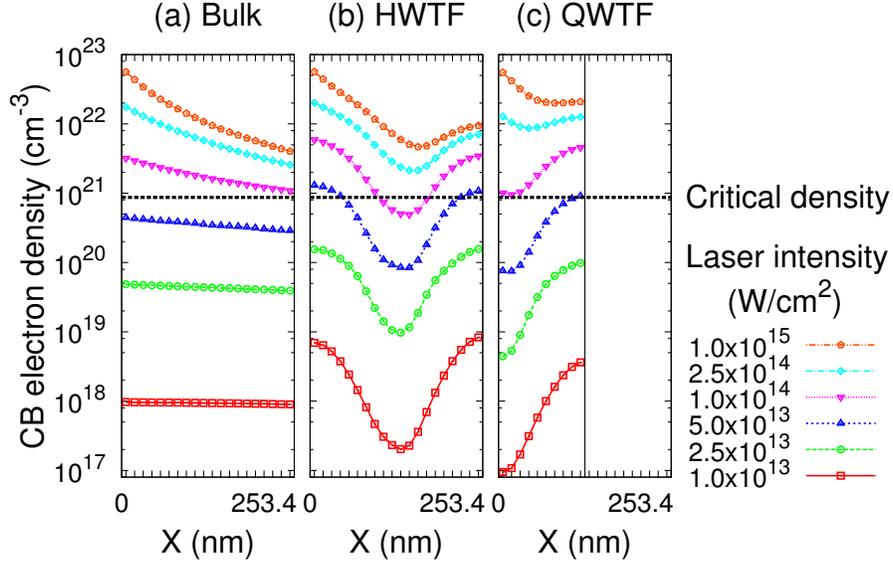}
\caption{\label{fig:CBEdensity} Spatial distribution of conduction band electrons in $\alpha$-quartz medium after its interaction with laser pulses of various laser intensities: (a) Bulk, (b) HWTF and (c) QWTF. $X$ refers to the distance from the sample surface at $X=0$: the bulk surface for bulk and the front surface for the films. The horizontal dashed line represents the critical density ($N_{\mathrm{cr}}$), the value of which is $8.7 \times 10^{20} \, \mathrm{cm^{-3}}$ for $\lambda_{0}=800 \, \mathrm{nm}$ and $m_{\mathrm{e}}^{\ast}=0.5 \, m_{\mathrm{e}}$.}
\end{figure*}

Figure \ref{fig:CBEdensity} shows the spatial distributions of CB electrons at various laser intensities after the pulse passes the medium. Since there is no interference in the bulk, the spatial distribution of the CB electrons in the bulk $\alpha$-quartz was relatively uniform inside the medium at low intensities (See Fig. \ref{fig:CBEdensity}(a)). As the intensity increases, the CB electrons were accumulated around the front surface and exceeded the critical density when $I_{0}=6 \times 10^{13} \, \mathrm{W/cm^{2}}$, at which the bulk reflectance increased only by $5 \, \%$.

A non-uniform spatial distribution of the CB electrons was observed for thin-film cases. For HWTF, more CB electrons were generated around the two surfaces than in the middle of HWTF, due to interference between the secondary waves generated from the two surfaces. With increases in intensity, the CB electrons exceeded the critical density at the two surfaces for a given intensity. The CB electrons generated around the front and rear surfaces change the interference condition in a way that destructively contributes to the reflectance. The change of the interference condition increases the reflectance because the destructive interference effect is reduced. Since we consider a laser pulse with a finite pulse duration, the broad spectrum of the laser pulse can affect the interference condition and thus the reflectance, for example, in the case of HWTF it can give rise to non-zero values of the CB electrons at the center of the film and also of reflectance at low intensities.

For QWTF, the CB electrons were dominantly generated around the rear surface and exceeded the critical density on increasing the laser intensity. As in the case of HWTF, the CB electrons above the critical density modify the interference condition. However, in the case of QWTF, the interference constructively contributes to the reflectance. This indicates that the change from an initial constructive interference condition can reduce the reflectance. As seen in Fig. \ref{fig:reflectance}, the reflectance for QWTF started to decrease at an intensity of $5 \times 10^{13} \, \mathrm{W/cm^{2}}$, at which the CB electrons reached the critical density on the rear surface, and had the minimum reflectance at an intensity of $1.7 \times 10^{14} \, \mathrm{W/cm^{2}}$, at which the CB electrons density reached up to $9 \times 10^{21} \, \mathrm{cm^{-3}}$. This value is much higher than the critical density. Therefore, it can be inferred that significant change in reflectance for QWTF occurs by the CB electrons above the critical density.

At higher intensities above $5 \times 10^{14} \, \mathrm{W/cm^{2}}$, more CB electrons were populated at the front surface in all the structures. Since a major part of the pulse was reflected from the front surface, interference between the secondary waves from the front and rear surfaces does not play an important role anymore and all the reflectances converged and rapidly increased to the reflectance of the over-dense plasma.

It should be noted that, at a given laser intensity, the films had a larger population of CB electrons than the bulk at the front and/or the rear surfaces. This implies that the optical response of the films might be more sensitive to changes in the laser intensity than the bulk.

\subsection{\label{ssec:absorption} Estimation of ablation threshold}

\begin{figure}
\includegraphics[width=80mm]{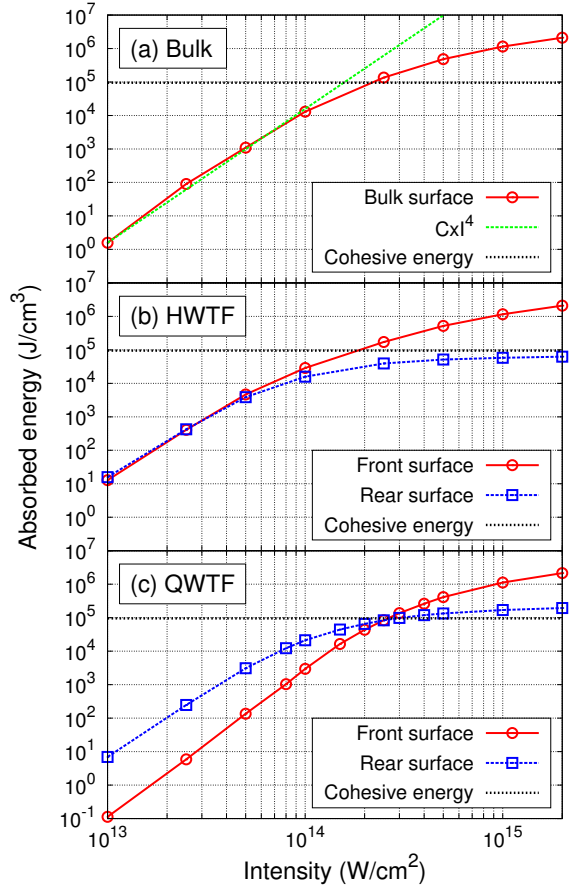}
\caption{\label{fig:absenergy} Absorbed energy at the surfaces as a function of laser intensity: (a) bulk, (b) HWTF and (c) QWTF. The horizontal dotted line represents the cohesive energy calculated by the LDA functional, $9.5 \times 10^{4} \, \mathrm{J/cm^{3}}$ \cite{Liu:1994}.}
\end{figure}

The CB electrons can cause a permanent structural damage of materials by transferring their kinetic energy to the lattice, in addition to changing the optical response. A knowledge of the damage threshold is thus equally important in designing an optical device such as a high-damage-threshold mirror. A criterion indicating the occurrence of damage is needed to establish a damage threshold. In this study, an energy criterion was used rather than the critical density as a criterion \cite{Penano:2005, Jia:2003, Chimier:2011}. Since we used a $20 \, \mathrm{fs}$ laser pulse, the typical damage type was laser-induced ablation; so the cohesive energy was adopted as a criterion for laser-induced ablation \cite{Chimier:2011}. For a consistent description, a cohesive energy value of $9.5 \times 10^{4} \, \mathrm{J/cm^{3}}$ obtained by the LDA functional rather than the experimental value of $8.2 \times 10^{4} \, \mathrm{J/cm^{3}}$ was used in the simulation. Note that the $15 \, \%$ difference in the cohesive energy did not severely influence our interpretations. The absorbed energy in the simulation was defined as the difference in total energies between before and after the laser pulse passes through a unit cell that was closest to the front surface and the rear surface. Figure \ref{fig:absenergy} shows the energies absorbed by the bulk and thin film samples at various laser intensities. For the bulk case, at laser intensities below $I_{0}=5 \times 10^{13} \, \mathrm{W/cm^{2}}$, the increase in the absorbed energy was proportional to $I_{0}^{4}$, as seen in Fig. \ref{fig:absenergy}(a) (see the dashed line), which is an evidence for four-photon absorption. The rate of increase in the absorbed energy became less as the intensity increased. This could be related to an increase in the reflectance, since a high reflectance implies a relatively low fraction of the transmitted wave, which in turn implies reduced availability of the wave for absorption, and hence a decrease in the energy absorbed by the medium.

The ablation threshold was determined from a cross point at which the absorbed energy and the cohesive energy intersected. The calculated ablation threshold at the bulk surface was $2.2 \times 10^{14} \, \mathrm{W/cm^{2}}$ and the corresponding fluence was $1.7 \, \mathrm{J/cm^{2}}$. This value is slightly lower than the experimental value of $2 \, \mathrm{J/cm^{2}}$ reported in Ref. \onlinecite{Xu:2007}. Although our simulation underestimated the band gap energy and considered only the multiphoton ionization for CB electron generation, the estimated ablation threshold showed a good agreement with the experimental value. The minor difference might have come from the underestimation in the band gap energy, which can be systematically modified by using a more elaborate functional that reproduces the experimental band gap energy. It should be mentioned that our estimated ablation threshold can be considered as a maximum operational intensity, which means the maximum intensity that the material can be exposed to without producing any ablation. This is because we assumed that the kinetic energy of the CB electrons was completely transferred to the lattice, while in reality some losses may actually be occurring, pushing the threshold higher up.

It is worth comparing the estimated threshold value with the one obtained by using another threshold criterion, i.e., the critical density. When the critical density was used, the threshold fluence for laser-induced ablation was $0.5 \, \mathrm{J/cm^{2}}$, which is significantly lower than that obtained from experiments. This may again be coming from the underestimated band gap energy used in the simulation. If the band gap energy becomes larger and closer to the experimental value, the electrons will not be easily excited from the VB to the CB. Therefore, a higher laser intensity would be needed to reach the critical density, resulting in an increase in ablation threshold fluence. However, the exact relation between the band gap energy and the generation of CB electrons needs to be investigated for an accurate evaluation of the critical density criterion.

The absorbed energies for the case of thin films showed interesting characteristics, depending on the thickness and the surface, whether front or rear. For HWTF, the absorbed energies at the front and rear surfaces had similar values at low laser intensities. However, the absorbed energy at the front surface was more than the rear surface at higher laser intensities. This can be understood from the CB electrons distributed around the front surface. The threshold for HWTF was $1.8 \times 10^{14} \, \mathrm{W/cm^{2}}$, which was slightly lower than that for the bulk, indicating that the HWTF is weaker than the bulk when it comes to the intense laser pulses. On the other hand, for QWTF, the absorbed energies at the front and rear surfaces showed a huge difference in the low laser intensity range, but the difference gradually decreased in the high intensity region. The calculated ablation thresholds for QWTF were $2.6 \times 10^{14} \, \mathrm{W/cm^{2}}$ for the front surface and $2.9 \times 10^{14} \, \mathrm{W/cm^{2}}$ for the rear surface. These results for the absorbed energy in the case of QWTF can be explained by the fact that at low laser intensities the CB electrons are mainly generated near the rear surface, but at high intensities the CB electrons near the front surface become dominant in the absorption process. Consequently, the ablation threshold is closely related to CB electrons generation and energy absorption by the CB electrons generated inside the medium.

\section{\label{sec:conclusion} Conclusion}

Through first-principles simulations, we have investigated the changes in the optical response of bulk and thin-film $\alpha$-quartz when irradiated with an intense ultrashort laser pulse. The generation of CB electrons in the medium has also been investigated in detail. The change in the reflectance with laser intensity was mainly attributed to the generation and spatial distribution of the CB electrons in the medium. The simulation studies performed for laser intensities in the range $10^{10} \, \mathrm{W/cm^{2}}$ to $2 \times 10^{15} \, \mathrm{W/cm^{2}}$ successfully described the transition (from a dielectric to plasma) property of the medium as well as the laser intensity required for this change in the optical properties. At low laser intensities, the interference effect between the secondary waves from the front and rear surfaces was the dominant process in the reflectance behavior of thin films. However, at high laser intensities, the CB electrons generated on the front surface played a dominant role and interference a minor role in changing the reflectance.

The energy absorbed by the CB electrons in the medium was used to estimate the laser-induced ablation threshold of $\alpha$-quartz materials. The theoretical estimation showed a good agreement with the experimental value, despite some limitations in the simulations, particularly, underestimation of the band gap energy. This limitation can be easily overcome by using more elaborate functionals; these will be discussed in a further study. The results obtained in our study can provide fundamental information on the parameters required for designing a high-performance optical coating structure, such as a high-damage-threshold and broadband multilayer coated mirror.

\begin{acknowledgments}

This work was supported by the Ministry of Trade, Industry and energy of Korea through the Infrastructure for femto-technology program supervised by the National IT Industry Promotion Agency. This work was also supported by the Grants-in-Aid for Scientific Research Nos. 23340113, 23104503, 21340073 and 21740303. Numerical calculations were partly performed on the K-Computer, Kobe, Japan, in early access stage.

\end{acknowledgments}

\end{document}